\documentclass[aps,prc,showpacs,twocolumn,preprintnumbers,floatfix,
showkeys,tightenlines]{revtex4}
\usepackage{amsmath,amssymb,amsfonts}
\usepackage{graphicx}
\usepackage{color}
\allowdisplaybreaks

\newcommand{\be}{\begin{equation}}
\newcommand{\ee}{\end{equation}}
\newcommand{\bea}{\begin{eqnarray}}
\newcommand{\eea}{\end{eqnarray}}
\newcommand{\bm}{\bibitem}

\newcommand{\ze}{\zeta}

\newcommand{\cz}{{\cal Z}}

\begin{document}



\title{ Warm alpha-nucleon matter }

\author{S. K. \surname{Samaddar}}
\email{santosh.samaddar@saha.ac.in}
\author{J. N. \surname{De}}
\email{jn.de@saha.ac.in}
\affiliation{
Saha Institute of Nuclear Physics, 1/AF Bidhannagar, Kolkata
{\sl 700064}, India} 


\begin{abstract} 
The properties of warm dilute alpha-nucleon matter are studied in a
variational approach in the Thomas-Fermi approximation starting from
an effective two-body nucleon-nucleon interaction. The equation of
state, symmetry energy, incompressibility of the said matter as well
as the $\alpha $ fraction are in consonance with those evaluated
from the virial approach that sets a bench-mark for such calculations 
at low densities.
\end{abstract}

\pacs{21.65.Cd, 21.65.Ef, 21.65.Mn, 24.10.Pa}

\keywords{ nuclear matter, alpha-clusters, Thomas-Fermi approach,
$S$-matrix} 

\maketitle

\section{Introduction}

Cold nuclear matter at subsaturation density as $\alpha $ matter
has been subjected to a critical study for some time
\cite{cla,car}. The aim is to understand the $\alpha $-clustering
near the surface of heavy nuclei or the putative dilute 
alpha-condensate in light 4n nuclei. In astrophysical context,
in following the evolution of the core-collapse supernovae,
these studies have been extended to the case of warm nuclear 
matter \cite{lamb}. The homogeneous low-density 
nuclear matter stabilizes as a mixture of nucleons and
nucleon-clusters. It has lower free energy compared to
that for nucleonic matter.  The cluster
composition is temperature and density dependent, with increasing
temperature or decreasing density, the population of heavier clusters
tends to diminish leading to a mixture of nucleons and light clusters
like d, t, $^{3}$He and $\alpha $ \cite{fri,pei}. 
The properties of the clusterized matter undergo a major change,
{\it e.g.}, 
 the incompressibility of clusterized nuclear matter is quite
smaller compared to that for homogeneous nucleon matter 
\cite{sam}. This directly influences  the collapse and bounce phase 
of the supernova matter. The symmetry energies of nuclear matter are
also affected significantly when matter gets clusterized \cite{hor,de}.
This has an important  role in a better understanding of neutrino-driven
energy transfer in supernova matter \cite{jan}. The symmetry energy also
influences the cluster composition in the crust  
of neutron stars and is thus
instrumental in shaping the details of their mass, cooling and structure
\cite{fuc}.

The equation of state (EOS) of warm dilute nuclear matter with only
light clusters upto $\alpha $ has recently been investigated in the
virial approach \cite{hor,con}; inclusion of heavier clusters has
also been made in the $S$-matrix (SM)
framework \cite{mal}. These
methods relate the calculations directly to the experimental observables
like the binding energies and the phase-shifts and thus, as such, are
model-independent. They are usually taken as {\it bench-mark } calculations
in the domain of low-density and high-temperature;  they are understood
to exhaust all the dynamical information concerning the strong
interactions in the medium.
For an 
interacting quantum  gas, the virial expansion, however, 
virtually ends at the second
order. Formulation of higher order virial coefficients are very
involved even at the formal level \cite{pai}, making it difficult to estimate
the domain of validity of the virial series truncated at the second
order. It may further be noted that
the density should be dilute enough so that 
the concept of asymptotic wave functions as inherent in the virial
expansion should be meaningful.

An alternate avenue could be  to bypass the virial
expansion altogether and take recourse to nucleation in the 
framework of the mean-field model with a suitably chosen effective
two-nucleon interaction that inherently takes an inclusive account
of the scattering effects. Unlike the virial ($S$-matrix) approach
which has direct contact with the experimental data,
this method has indirect contact but it can
be applied to relatively higher densities.
 With increasing density, a large number of
different fragment species would, however, be  formed that makes the numerical
calculation very lengthy. Before attempting any full-blown calculation,
it may then be worthwhile as a first step,
to take only $\alpha $-clustering in the nuclear matter and to 
examine  whether the model works in the low-density region where
the bench-mark calculations exist. The present work aims
towards that end.

For the study of the so-mentioned $\alpha $-nucleon ($\alpha $N) matter,
we have chosen the Thomas-Fermi prescription for the mean-field model
and the finite range, momentum and density dependent modified
Seyler-Blanchard (SBM) effective interaction \cite{de1}. The properties
we explore include the EOS of the $\alpha $N matter, its symmetry energy,
incompressibility, $\alpha $ concentration etc. In Sec. II, the theoretical
framework for the mean-field and the $S$-matrix approach
is presented. Sec. III contains the results and discussions.
Concluding remarks are given in Sec. IV.

\section{Theoretical framework}
 
 Given an effective two-nucleon interaction, the properties of the
$\alpha $N matter can be evaluated by exploiting the occupation 
functions of the n, p and alphas obtained from minimization
of the thermodynamic potential of the system. In Sec. II~A, some details
of the effective interaction used are given. In Sec II~B,
theoretical formulation for obtaining the occupation functions from 
the Thomas-Fermi (TF) approximation
is presented. In Sec. II~C, expressions for various observables
explored are given. In Sec.~II~D, a brief outline of the $S$-matrix approach
is made.

\begin{table}
\caption{ The parameters of the effective interaction (in MeV fm units)}
\begin{ruledtabular}
\begin{tabular}{cccccc}
$C_l$&  $C_u$& $a$& $b$& $d$& $\kappa $\\
\hline
291.7& 910.6& 0.6199& 928.2& 0.879& 1/6\\
\end{tabular}
\end{ruledtabular}
\end{table}

\subsection{The effective interaction}

The form of the SBM effective interaction $v$ is

\begin{eqnarray}
v(r,p,\rho )&=&C_{l,u}\left [v_1(r,p)+v_2(r,\rho )\right ], \nonumber \\
v_1&=&-(1-\frac{p^2}{b^2})f({\bf r_1,  r_2}), \nonumber \\
v_2&=&d^2\left [ \rho (r_1) + \rho (r_2) \right ]^\kappa f({\bf r_1, r_2}), 
\end{eqnarray}
with 
\begin{eqnarray}
f({\bf r_1, r_2})&=& \frac{e^{-|{\bf r_1-r_2 }|/a}}{|{\bf r_1-r_2 }/a}.
\end{eqnarray}
The subscripts $l$ and $u$ to the interaction strength $C$ refer to 
like-pair (nn or pp) and unlike-pair (np) interactions, respectively.
The range of the effective interaction is given by $a, ~b$ is the measure
of the strength of the momentum dependence of the interaction. The
relative separation of the interacting nucleons is ${\bf r=r_1-r_2 }$
and the relative momentum is ${\bf p=p_1-p_2 }$; $d$ and $\kappa$ are
the two parameters governing  the strength of the density dependence
and $\rho (r_1)$ and $\rho (r_2)$ are the nucleon densities at the sites
of the two interacting nucleons. The potential parameters are given
in Table~I; the details for the determination of these 
parameters are given in \cite{de1}. \\
The incompressibility $K_\infty$ of 
symmetric nucleon matter is mostly governed by the 
parameter $\kappa$; for the potential set we have chosen, the value
of $K_\infty$=238 MeV.

The equation of state of symmetric nuclear matter calculated with the
SBM interaction is seen to agree extremely well \cite{uma} with
that obtained in a variational approach by Friedman and Pandharipande
\cite{fri} with $v_{14}$+TNI interaction. This interaction
also reproduces quite well the  binding energies,
rms charge radii, charge distributions
and the giant monopole resonance energies for a host of even-even nuclei
ranging from $^{16}$O to very heavy systems \cite{de1}. Interactions
of this type has been used with great success by Myers and Swiatecki
\cite{mye} in the context of nuclear mass formula.

\subsection{The occupation functions}

 The self-consistent occupation probabilities of nucleons and alphas
in $\alpha $N matter at temperature $T$  are
obtained  in the TF approximation 
by minimizing the thermodynamic potential of the system
\begin{eqnarray}
\Omega = E-TS-\sum_\tau \mu_\tau N_\tau -\mu_\alpha N_\alpha .
\end{eqnarray}
Here $\tau $ represents the isospin index (n,p). The quantities $E,S,
\mu_\tau, \mu_\alpha, N_\tau $ and $N_\alpha $ are the total internal
energy, entropy, nucleon chemical potentials, $\alpha $ chemical potential,
free nucleon numbers and the number 
of $\alpha $ particles, respectively, in
the system. Chemical equilibrium in the system ensures 
\begin{eqnarray}
\mu_\alpha = 2(\mu_n +\mu_p ).
\end{eqnarray}
The total energy of the $\alpha $N matter in TF approximation is
\begin{eqnarray}
&&E= \sum_\tau\Biggl  \{\int d{\bf r_1} d{\bf p_1} \frac{p_1^2}{2m_\tau } 
\tilde n_\tau ({\bf p_1}) +\frac{1}{2} \int d{\bf r_1}d{\bf p_1}
d{\bf r_2}d{\bf p_2} \nonumber  \\
&&\times [v_1(|{\bf r_1}-{\bf r_2}|, |{\bf p_1}
-{\bf p_2}|)+v_2(|{\bf r_1}-{\bf r_2}|,2\rho)]
[C_l\tilde n_\tau ({\bf p_2}) \nonumber \\
&&+C_u \tilde n_{-\tau }({\bf p_2})] \tilde n_\tau ({\bf p_1}) 
+\frac{1}{2}(C_l +C_u )\int d{\bf r_1} d{\bf p_1} d{\bf r_2}
d{\bf p_2} \nonumber \\
&&\times \tilde n_\tau ({\bf p_1}) \tilde n_\alpha ({\bf p_2})
\int_{V_\alpha} d{\bf r}\int d{\bf p_i^\alpha }\tilde n_i^\alpha 
({\bf p_i^\alpha}) [v_1({\bf R^{\prime }},|{\bf p_1}-({\bf p_i^\alpha}
\nonumber \\
&&+{\bf p_2})|) + v_2({\bf R^{\prime }}, (\sum_{\tau^{\prime }}
\rho_{\tau^{\prime}}+\rho_i^\alpha))] \Biggr \} 
\nonumber \\
&&+\int d{\bf r_1}d{\bf p_1} 
\frac{p_1^2}{2m_\alpha}\tilde n_\alpha ({\bf p_1}) \nonumber \\
&& +(C_l+C_u)\int d{\bf r_1}d{\bf p_1} d{\bf r_2} d{\bf p_2}
\tilde n_\alpha ({\bf p_1})\tilde n_\alpha ({\bf p_2}) \nonumber \\
&& \times \int_{V_\alpha } d{\bf r} d{\bf r^\prime }
\int d{\bf p_i^\alpha}
d{\bf p_i^{\alpha ^{\prime}}}\tilde n_i^\alpha ({\bf p_i^\alpha})
\tilde n_i^{\alpha^{\prime}}({\bf p_i^{\alpha^{\prime}}}) \nonumber \\
&& \times [v_1({\bf R^{\prime}}, |({\bf p_1}+{\bf p_i^\alpha })
-({\bf p_2}+{\bf p_i^{\alpha^{\prime}}}|)+v_2({\bf R^{\prime}},
2\rho_i^\alpha)] \nonumber \\
&& -N_\alpha B_\alpha
\end{eqnarray}
 In Eq.~(5), $m_\tau $ and $m_\alpha $ are the nucleon and 
$\alpha $ masses, the first and fourth terms correspond to the kinetic energy
of the nucleons and alphas, the second and the fifth terms refer
to the interaction energy among nucleons and among alphas, 
respectively and the third term is the  interaction energy
between nucleons and alphas.
The various space coordinates occurring in the third and fifth terms are shown 
in Figs.~1 and 2, respectively.  

\begin{figure}
\includegraphics[width=1.0\columnwidth,angle=0,clip=true]{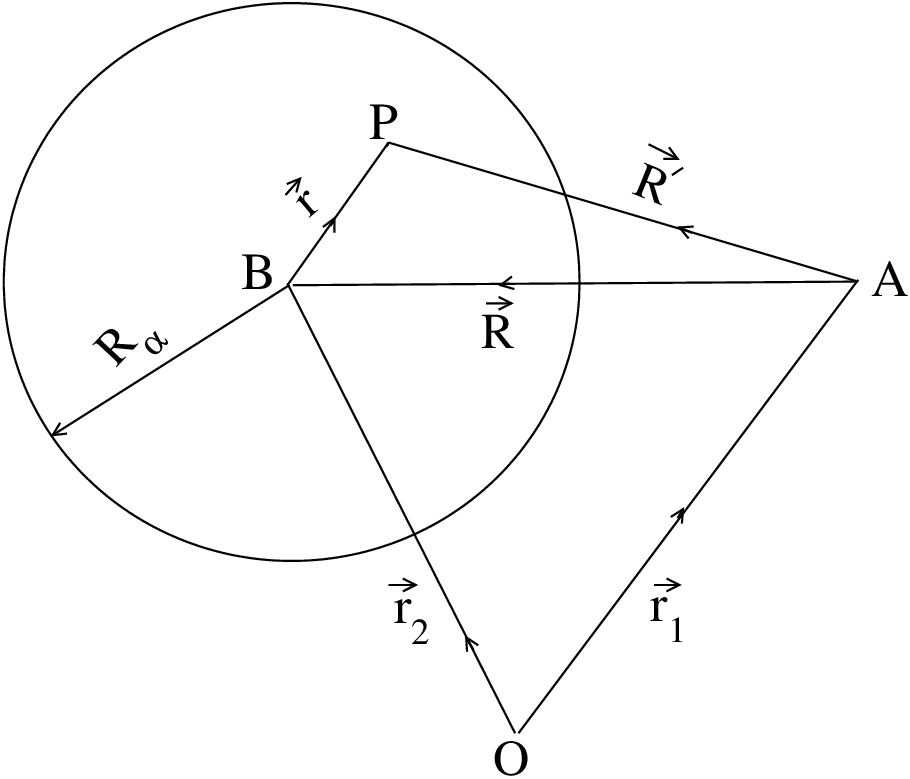}
\caption{ Space coordinates shown
for nucleon (located
at $A$) and alpha (with center at $B$) configuration. The origin 
of the coordinate system is at $O$
and $P$ is any arbitrary point within alpha.}
\end{figure}

\begin{figure}
\includegraphics[width=1.0\columnwidth,angle=0,clip=true]{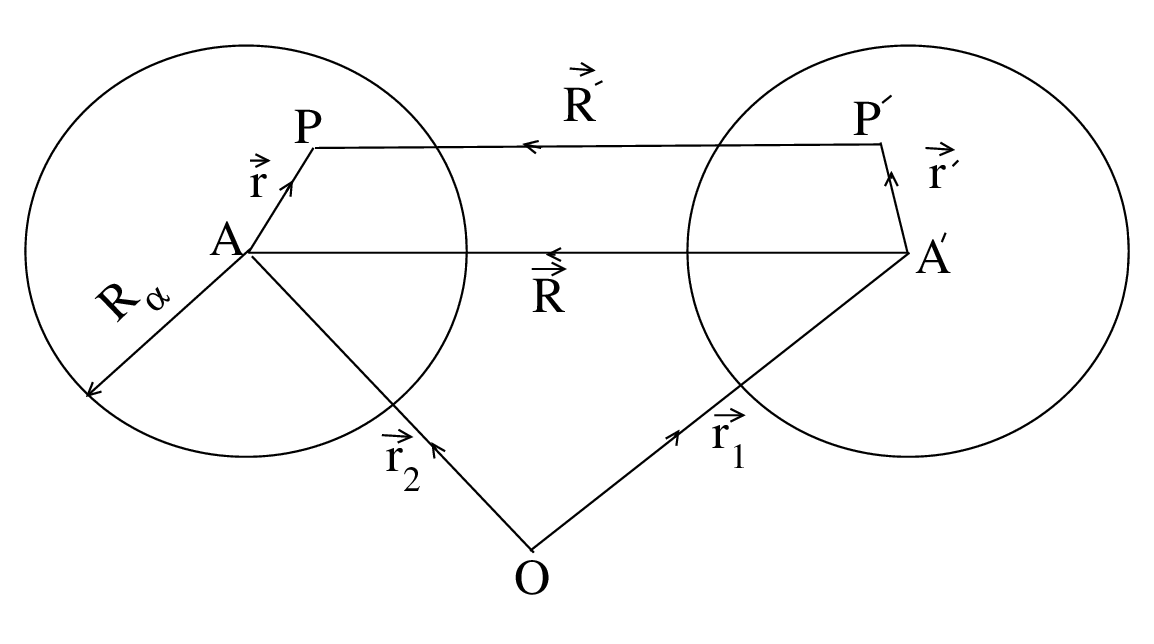}
\caption{ Space coordinates shown for alpha-alpha configuration with
 $O$ as the origin
of the coordinate system. $P$ and
$P^{\prime}$ are arbitrary points within the alphas 
with $A$ and $A^{\prime}$ as their centers.}
\end{figure}

These terms are evaluated in the single-folding
and double-folding models. The last term is the binding
energy contribution from the $\alpha $ particles. Here $\tilde n_\tau =
\frac{2}{h^3}n_\tau, \tilde n_\alpha =\frac{1}{h^3}n_\alpha $ where
$n_\tau $ and $n_\alpha $ are the occupation probabilities for 
nucleons and alphas, respectively. Similarly, 
$\tilde n_i^{\alpha }({\bf p_i^{\alpha }})=\frac{2}{h^3}
n_i^{\alpha }({\bf p_i^{\alpha }})$ represents the occupation
probability of the constituent nucleons in the $\alpha $ particle
and ${\bf p_i^{\alpha }}$ is their intrinsic momentum inside 
the $\alpha $. The space coordinates do not enter in the occupation
functions $\tilde n_\tau $ and $\tilde n_\alpha $ as the system 
is infinite. For simplicity, the $\alpha $ particles are taken to 
be uniform nuclear drops with a sharp surface and hence the space
coordinates do not also occur in $\tilde n_i^{\alpha }$. The
notation $\int_{V_{\alpha }}$ refers to configuration integral
over the volume of $\alpha $. The integral over $\tilde {\bf p_i^\alpha }$
is over the Fermi sphere of the nucleon momenta inside the $\alpha $
particles. Since alphas are difficult to excite (the first excited
state in $\alpha $ is $\sim $ 20 MeV), they are
taken to be in their ground states.
All the other integrals are over the 
entire configuration or momentum  space unless otherwise specified. 
It then follows that 
\begin{eqnarray} 
\int \tilde n_\tau ({\bf p})d{\bf p}= N_\tau/V=\rho_\tau , \nonumber \\
\int \tilde n_\alpha ({\bf p})d{\bf p} =N_\alpha /V=\rho_\alpha ,
\nonumber \\
\int \tilde n_i^{\alpha }({\bf p}) d{\bf p}=4/V_\alpha =\rho_i^{\alpha } ,
\end{eqnarray} 
where $V$ is the volume of the $\alpha$N system and 
$V_\alpha =\frac{4}{3}\pi R_\alpha ^3 $, with $R_\alpha $ as 
the sharp-surface radius of the $\alpha $ drop taken to be 2.16 fm
obtained from experimental rms charge-radius of $\alpha $; 
$\rho_i^\alpha $ is the density of the constituent nucleons of
the $\alpha $ particles.
The total
baryon density $\rho_b $ is given by $\rho_b=\rho +4\rho_\alpha $ 
where $\rho =\sum_\tau \rho_\tau $ is the density of the free
nucleons and $\rho_\alpha $ is the $\alpha $-particle density.

The total entropy of the $\alpha $N system is 
\begin{eqnarray} 
S=\sum_\tau S_\tau +S_\alpha , 
\end{eqnarray} 
where in the Landau quasi-particle approximation, 
\begin{eqnarray} 
S_\tau &=&\frac{2}{h^3} \int \bigl [n_\tau ({\bf p}) \ln n_\tau ({\bf p})
\nonumber \\
&&+(1-n_\tau ({\bf p}))\ln (1-n_\tau ({\bf p})) \bigr ] d{\bf r}d{\bf p},
\end{eqnarray} 
and
\begin{eqnarray} 
S_\alpha & =& \frac{1}{h^3} \int \bigl 
[n_\alpha ({\bf p}) \ln n_\alpha ({\bf p}) \nonumber \\
&& -(1+n_\alpha ({\bf p}))\ln (1+n_\alpha ({\bf p})) \bigr ] d{\bf r}d{\bf p}.
\end{eqnarray} 
Minimization of $\Omega $ with respect to $n_\tau $ and $n_\alpha $,
remembering that $\delta n_\tau ({\bf p})$ and $\delta n_\alpha 
({\bf p})$ are separately arbitrary over the whole phase space,
at the end yields 
\begin{eqnarray} 
&&\frac{p_1^2}{2m_\tau }+\int d{\bf r_2}d{\bf p_2}\Bigl \{v_1(|{\bf r_1}
-{\bf r_2}|,|{\bf p_1}-{\bf p_2}|) \nonumber \\
&&+v_2(|{\bf r_1}-{\bf r_2}|,2\rho )\Bigr \} 
[C_l\tilde n_\tau ({\bf p_2})+C_u\tilde n_{-\tau }({\bf p_2})] \nonumber \\
&& +\kappa d^2 (2\rho ) ^{\kappa -1}\sum_{\tau ^{\prime }}\int 
d{\bf p_1^{\prime }}d{\bf r_2}d{\bf p_2} \nonumber \\
&&\times [C_l \tilde n_{\tau^{\prime }}
({\bf p_2})+C_u\tilde n_{-\tau^{\prime }}
({\bf p_2})]\tilde n_{\tau^{\prime }}
({\bf p_1}^{\prime })f({\bf r_1},{\bf r_2}) \nonumber \\
&&+\frac{1}{2}
(C_l+C_u)\int d{\bf r_2}d{\bf p_2}\tilde n_\alpha ({\bf p_2}) \nonumber \\
&& \times \int d{\bf r}d{\bf p_i^\alpha }
\tilde n_i^{\alpha }({\bf p_i^{\alpha }}) 
 \bigl \{v_1({\bf R^{\prime}},|{\bf p_1}-({\bf p_i^{\alpha }}+ 
{\bf p_2})|) \nonumber \\
&& +v_2({\bf R^{\prime }},(\rho +\rho_i^{\alpha } )) \bigr \} 
+\frac{1}{4}(C_l+C_u)\kappa d^2(\rho +\rho_i^\alpha )^{\kappa -1} \nonumber \\
&& \times \sum_{\tau^{\prime }}\int d{\bf p_1^\prime }d{\bf p_2}
\tilde n_{\tau^{\prime }}({\bf p_1^\prime })\tilde n_{\alpha }({\bf
p_2 })\rho_i^\alpha \nonumber \\
&&\times \int d{\bf r_2}\int_{V_{\alpha }}d{\bf r} 
\frac{e^{-|{\bf R ^{\prime }}|/a }}{|{\bf R ^{\prime }}|/a } \nonumber \\
&& +T\bigl [\ln n_{\tau }({\bf p_1 })
-\ln (1-n_{\tau }({\bf p_1 }))\bigr ] 
 -\mu_{\tau}=0 ,
\end{eqnarray} 
and 
\begin{eqnarray} 
&& \frac{p_1^2}{2m_{\alpha }}+2(C_l+C_u)\int d{\bf r_2}d{\bf p_2}
\tilde n_{\alpha}({\bf p_2}) \nonumber  \\ 
&&\times \int d{\bf r}d{\bf p_i^{\alpha }}d{\bf r^{\prime }}d{\bf p_i^
{\alpha \prime }}
\tilde n_i^\alpha ({\bf p_i^\alpha })\tilde n_i^{\alpha \prime }
({\bf p_i^{\alpha \prime }}) \nonumber \\
&&\times \Bigl \{v_1({\bf R^\prime },|({\bf p_1+p_i^\alpha })-({\bf p_2+p_i^{\alpha
\prime }})|)+v_2({\bf R^\prime },2\rho_i^\alpha ) \Bigr \} \nonumber \\
&& +\frac{1}{2}(C_l+C_u)\sum_\tau \int d{\bf p_2}
d{\bf p_i^\alpha }\tilde n_\tau
({\bf p_2}) \tilde n_i^\alpha ({\bf p_i^\alpha }) \nonumber \\
&& \times \int d{\bf r_2}\int_{V_\alpha }d{\bf r}
 \Bigl \{v_1({\bf R^\prime },|{\bf p_2}-
({\bf p_1}+{\bf p_i^\alpha })|) \nonumber \\
&& +v_2({\bf R^\prime },
\rho +\rho_i^\alpha ) \Bigr \} +T \bigl [\ln n_\alpha ({\bf p_1})-\ln (1+
n_\alpha ({\bf p_1 })) \bigr ] \nonumber \\
&& -(\mu_\alpha +B_\alpha )=0. 
\end{eqnarray} 
Without any loss of generality, ${\bf r_1}$ can be set equal to 
zero in Eqs.~(10) and (11).
The single-particle occupation functions $n_\tau (p)$ and $n_\alpha (p)$
for nucleons and alphas are determined from Eqs.~ (10) and (11),
respectively. 
 Eq.~(10), after some algebraic manipulations can be written as
\begin{eqnarray}
&& \frac{p_1^2}{2m_\tau}+V_\tau^0+p_1^2V_\tau^1+V_\tau^2 \nonumber \\
&& + T \bigl [\ln n_\tau ({\bf p_1})
-\ln (1-n_\tau ({\bf p_1})) \bigr ]-\mu_\tau =0.
\end{eqnarray} 
The momentum-dependent nucleon single-particle potential 
$V_\tau (p) $ is given by 
\begin{eqnarray} 
V_\tau (p)=V_\tau^0+p^2V_\tau^1 ,
\end{eqnarray} 
  where $V_\tau^0 $ is the momentum-independent part. Eq.~(12) leads
to
\begin{eqnarray} 
n_\tau (p)=\left [1+exp \left \{ \left ( \frac{p^2}{2m_\tau^*}
+V_\tau^0+V_\tau^2-\mu_\tau \right )/T \right \} \right ]^{-1},
\end{eqnarray} 
where $m_\tau^* $ is the nucleon effective mass,
\begin{eqnarray} 
m_\tau^*=\left [\frac{1}{m_\tau}+2V_\tau^1 \right ]^{-1},
\end{eqnarray} 
and $V_\tau^2 $ is the rearrangement potential coming from the
density dependence of the interaction. Similarly Eq.~(11) can
be written as 
\begin{eqnarray} 
&& \frac{p_1^2}{2m_\alpha}+V_\alpha^0+p_1^2V_\alpha^1 
 + T \bigl [\ln n_\alpha ({\bf p_1})
-\ln (1+n_\alpha ({\bf p_1})) \bigr ] \nonumber \\
&& -(\mu_\alpha +B_\alpha ) =0,
\end{eqnarray} 
which yields
\begin{eqnarray} 
n_\alpha (p)=\left  [exp \left ( \left \{ \frac{p^2}{2m_\alpha^*}
+V_\alpha^0-(\mu_\alpha +B_\alpha  )\right \}/T  \right )
-1 \right ]^{-1}
\end{eqnarray} 
where 
\begin{eqnarray} 
m_\alpha^*=\left [\frac{1}{m_\alpha}+2V_\alpha^1 \right ]^{-1},
\end{eqnarray} 
is the $\alpha $  effective mass. $V_\alpha ^0 $ is the momentum-independent
part of the $\alpha $-single particle potential $V_\alpha 
(=V_\alpha ^0 +p^2V_\alpha ^1) $ in the system.
The nucleon and $\alpha $ masses
are renormalized due to the momentum dependence in the interaction.

The expressions for $V_\tau^0$
can be arrived at as,
\begin{eqnarray} 
&& V_\tau^0= -4\pi a^3\left \{ 1-d^2(2\rho )^\kappa \right \}
(C_l\rho_\tau +C_u\rho_{-\tau }) \nonumber \\
&& +\frac{16\pi^2a^3}{b^2h^3} \biggl [C_l(2m_\tau^*T)^{5/2}J_{3/2}
(\eta_\tau )+C_u(2m_{-\tau }^*T)^{5/2} \nonumber \\
&& \times J_{3/2}(\eta_{-\tau }) \biggr ] 
 +\frac{1}{4}I(C_l+C_u)\rho_\alpha \rho_\alpha^i 
\biggl [ \frac{<p_\alpha^2>+<(p_i^\alpha )^2>}{b^2} \nonumber \\
&& +d^2(\rho
+\rho_i^\alpha )^\kappa -1 \biggr ].
\end{eqnarray} 
The first two terms come from the  interaction
between free nucleons, the last
term originates from the presence of alphas.
In Eq.~(19), $I$ is the six-dimensional integral (see Fig.~1)
\begin{eqnarray} 
I=\int_{V_\alpha} d{\bf r} \int d{\bf R} \frac{e^{-|{\bf r}+{\bf R}|/a}}
{|{\bf r}+{\bf R}|/a}.
\end{eqnarray} 
 This integral can be evaluated analytically.
The quantity $<p_\alpha^2>$ is the mean squared value of the 
$\alpha$ momentum in $\alpha$N matter and $<(p_i^\alpha)^2>$ 
is the mean squared value of the constituent nucleon momentum inside
the $\alpha$. The value of $<p_\alpha^2>$ is
\begin{eqnarray} 
<p_\alpha^2>&=&(2m_\alpha^*T)B_{3/2}(\eta_\alpha)/B_{1/2}(\eta_\alpha)
\nonumber \\
&&\simeq 3m_\alpha^*T,
\end{eqnarray} 
and
\begin{eqnarray} 
<(p_i^\alpha)^2>\simeq \frac{3}{5}(P_F^\alpha)^2
\end{eqnarray} 
where $P_F^\alpha$ is the value 
of the zero-temperature nucleon Fermi momentum inside 
$\alpha$, taken to be 220.5 MeV/c, consistent with the $\alpha$
sharp surface radius. The $J_k(\eta )$ and $B_k(\eta )$ are the
Fermi and Bose integrals, 
\begin{eqnarray} 
J_k(\eta )=\int_0^\infty \frac{x^k~dx}{e^{(x-\eta)}+1},
\end{eqnarray} 
and
\begin{eqnarray} 
B_k(\eta )=\int_0^\infty \frac{x^k~dx}{e^{(x-\eta)}-1},
\end{eqnarray} 
with 
\begin{eqnarray} 
&& \eta_\tau =(\mu_\tau -V_\tau^0 -V_\tau^2 )/T, \nonumber \\
&& \eta_\alpha =(\mu_\alpha +B_\alpha -V_\alpha^0 )/T.
\end{eqnarray} 
 The expressions for $V_\tau^1 ,
V_\tau^2$, $V_\alpha^0$ and $V_\alpha^1 $ 
are given as
\begin{eqnarray} 
 V_\tau^1=\frac{4\pi a^3}{b^2}[C_l\rho_\tau +C_u \rho_{-\tau }]
+\frac{1}{4}I(C_l+C_u) \frac{\rho_\alpha \rho_\alpha^i }{b^2},
\end{eqnarray} 
\begin{eqnarray} 
&& V_\tau^2= 4 \pi a^3 \kappa d^2(2\rho )^{\kappa -1} \sum_{\tau \prime }
[C_l\rho_{\tau ^\prime }+C_u \rho_{-\tau ^\prime }]\rho_{\tau ^\prime }
\nonumber \\
&& +\frac{1}{4}I(C_l+C_u)\kappa d^2(\rho+\rho_i^\alpha )^{\kappa -1}
\rho_i^\alpha \rho_\alpha \rho, 
\end{eqnarray} 
\begin{eqnarray}
&& V_\alpha^0 = \frac{1}{4}(C_l+C_u)\rho_i^\alpha \Biggl \{2\rho_i^\alpha
\rho_\alpha I_\alpha \bigl [d^2(2\rho_i^\alpha )^\kappa -1 \nonumber \\
&& +\frac{3m_\alpha^*T+\frac{6}{5}(P_F^\alpha )^2}{b^2} \bigr ] 
+I \bigl [\rho \bigl \{ d^2(\rho+\rho_i^\alpha )^\kappa -1 \nonumber \\
&& +\frac{3}{5}\frac{(P_F^\alpha )^2 }{b^2} \bigr \}
+\sum_\tau \frac{ 4\pi (2m_\tau^*T)^{5/2} J_{3/2}(\eta_\tau )}
{h^3b^2} \bigr ] \Biggr \},
\end{eqnarray} 
and
\begin{eqnarray} 
V_\alpha^1=\frac{1}{4}(C_l+C_u)\rho_i^\alpha \bigl \{2\rho_i^\alpha 
\rho_\alpha I_\alpha +I\rho \bigr \}/b^2.
\end{eqnarray} 
In both $V_\tau^1 $ and $V_\tau^2 $, the last term stems from
the $\alpha $-N interaction. The effective nucleon mass in pure
nucleonic matter thus gets modified due to clusterization.
The integral $I_\alpha $ occurring in Eqs.~ (28) and 
(29) is a nine-dimensional
integral (see Fig.~2),
\begin{eqnarray} 
I_\alpha=\int_{V_\alpha }d{\bf r }\int_{V_\alpha }d{\bf r^\prime }
\int d{\bf R} \frac{e^{-|{\bf R}+{\bf r}-{\bf r^\prime }|/a}}
{|{\bf R}+{\bf r}-{\bf r^\prime }|/a},
\end{eqnarray} 
which can be evaluated numerically.
If the alphas do not interpenetrate, the integral over ${\bf R }$
excludes the $\alpha $ volumes.

\subsection{Expressions for observables in TF approximation}

{\bf i) Energy per baryon :}
The energy per baryon $e_b$ of the $\alpha$N matter can be calculated
from Eq.~(5). It can be split into the following form,
\begin{eqnarray} 
e_b=e_{NN}+e_{\alpha N}+e_{\alpha \alpha }.
\end{eqnarray} 
Here $e_{NN} $ comes from the kinetic energy of the free nucleons 
and the interactions among them, $e_{\alpha N }$ arises from the
interaction among the free nucleons and the alphas and 
$e_{\alpha \alpha }$ stems from the kinetic energy of the
alphas and the interaction among themselves. The expressions for
them are
\begin{eqnarray} 
 e_{NN}&=&\frac{1}{\rho_b}\sum_\tau \rho_\tau \bigl [TJ_{3/2}(\eta_\tau)/
J_{1/2}(\eta_\tau ) \{ 1-m_\tau^*V_\tau^1 \} \nonumber \\
&&+\frac{1}{2}V_\tau^0 \bigr ],
\end{eqnarray} 
\begin{eqnarray} 
&& e_{\alpha N}= \frac{1}{4\rho_b}(C_l+C_u)I\rho_\alpha \rho_\alpha^i
\Bigl [\Bigl \{\frac{3m_\alpha^*T+3/5(P_F^\alpha)^2}{b^2} \nonumber \\
&& -1+d^2(\rho +\rho_i^\alpha )^\kappa \Bigr \} \rho \nonumber \\
&& +\frac{1}{b^2}\sum_\tau \bigl (\frac{4\pi }{h^3}(2m_\tau^*T)^{5/2}
J_{3/2}(\eta_\tau ) \bigr ) \Bigr ],
\end{eqnarray} 
and
\begin{eqnarray} 
&& e_{\alpha \alpha }=\frac{1}{\rho_b} \Bigl [\frac{\pi }{m_\alpha h^3}
(2m_\alpha^*T)^{5/2}B_{3/2}(\eta_\alpha ) \nonumber \\
&& + \frac{1}{4}(C_l+C_u)I_\alpha \rho_\alpha^2 (\rho_i^\alpha )^2
\Bigl \{ d^2(2\rho_i^\alpha )^\kappa -1 
\nonumber \\
&& +\frac{6m_\alpha^*T}{b^2}
 +\frac{6}{5}\frac {(P_F^\alpha )^2}{b^2} \Bigr \} \Bigr ].
\end{eqnarray} 
In the above equations, as stated earlier, $\rho_b$ 
(=$\rho +4\rho_\alpha $) corresponds to the total baryon density,
$\rho $ and $\rho_\alpha $ are the free nucleon and $\alpha $
densities, respectively, in the $\alpha$N system. \\

{\bf ii) Entropy per baryon:} 
The entropy per baryon $s_b$ can be evaluated using Eqs.~(8) and (9). It is
additive and can be written as 
\begin{eqnarray} 
s_b=s_N+s_\alpha ,
\end{eqnarray} 
where $s_N$ and $s_\alpha $ are the contributions to entropy
from free nucleons and alphas respectively. Their expressions
reduce to 
\begin{eqnarray} 
s_N=\frac{1}{\rho_b }\sum_\tau \rho_\tau \Bigl 
[\frac{5}{3} J_{3/2}(\eta_\tau )
/J_{1/2}(\eta_\tau )-\eta_\tau \Bigr ],
\end{eqnarray} 
and 
\begin{eqnarray} 
s_\alpha =\frac{\rho_\alpha }{\rho_b } 
\Bigl [\frac{5}{3} B_{3/2}(\eta_\alpha )
/B_{1/2}(\eta_\alpha )-\eta_\alpha \Bigr ].
\end{eqnarray} 

\vskip 0.5cm

{\bf iii) Pressure of $\alpha $N matter:}
Once the energy and entropy of the composite system are known,
the pressure can be calculated from the Gibbs-Duhem thermodynamic
identity,
\begin{eqnarray} 
P=\sum_\tau \rho_\tau \mu_\tau +\rho_\alpha \mu_\alpha -f_b\rho_b,
\end{eqnarray} 
where $f_b$ is the free energy per baryon, $f_b=e_b-Ts_b $. \\

{\bf iv) Incompressibility and the symmetry coefficients:} 
The incompressibility
$K$ can be computed from the derivative of pressure
\begin{eqnarray} 
K=9\frac{dP}{d\rho}.
\end{eqnarray} 
The symmetry free energy and symmetry energy coefficients $C_F$ and 
$C_E$ are calculated from 
\begin{eqnarray} 
C_F=\frac{1}{2}\Bigl ( \frac{\partial^2 f_b}{\partial X^2} \Bigr )_{X=0},
\end{eqnarray} 
\begin{eqnarray} 
C_E=\frac{1}{2}\Bigl ( \frac{\partial^2 e_b}{\partial X^2} \Bigr )_{X=0},
\end{eqnarray} 
where $X$ is the neutron-proton asymmetry of the $\alpha $N system.
It is given as $X=(\rho_b^n-\rho_b^p)/\rho_b $, where $\rho_b^n$
and $\rho_b^p $ are the total neutron and proton
density, respectively.

\vskip 1.0cm

\subsection {The $S$-matrix approach}

The relevant key elements of the $S$-matrix framework \cite{das}
as applied in the context of dilute nuclear matter \cite{mal,de}
are outlined in brief below.

The grand partition function of an interacting infinite system of
neutrons and protons can be written as 
\be
\cz = \sum_{Z,N=0}^\infty (\ze_p)^Z (\ze_n)^N 
\, {\rm Tr}_{Z,N}\, e^{-\beta H}~.
\ee
where $\zeta_p =e^{\beta \mu_p}$ and $\zeta_n=e^{\beta \mu_n}$
are the elementary fugacities with $\beta =1/T$ and $\mu $'s are
the nucleonic chemical potentials. Here $H$ is the total Hamiltonian
of the system and the trace $Tr_{Z,N}$ is taken over states of
$Z$ protons and $N$ neutrons. The partition function can be split
into two types of terms \cite{das}
\be
\ln \cz =\ln {\cz}_{part}^{(0)} + \ln {\cz}_{scat}~.
\ee
The first term on the right hand side corresponds to contributions
from stable single-particle states of clusters of different sizes
including free nucleons formed in the system; the second term refers
to all possible scattering states. The superscript (0) indicates that
the clusters behave as an ideal quantum gas. In general, 
$\ln {\cz}_{part}^{(0)} $ contains contributions from the ground states
as well as the particle-stable excited states of all the clusters. 
The scattering term $\ln {\cz}_{scat}$ may be written as a sum
of scattering contributions from a set of channels, 
each set having total proton
number $Z_t$ and neutron number $N_t$. Since our interest in the
present work is focused on $\alpha $N matter, in $\ln {\cz}_{part}^{(0)}$,
we include only the nucleons and the ground state of $\alpha $; similarly
in $\ln {\cz}_{scat}$, only the scattering channels $NN, \alpha N $ and
$\alpha \alpha $ are considered, so that  
\be
\ln {\cz}_{scat}~=~ \ln {\cz}_{NN}
+\ln {\cz}_{\alpha N}+\ln {\cz}_{\alpha \alpha}.
\ee
Each of the terms in Eq.~(44) can be expanded in the respective virial
coefficients. Expansion upto the second-order coefficients are only 
considered. They are written as energy integrals of the relevant 
phase-shifts \cite{hor,sam}.  
 The partition function can then be written explicitly as
\bea
&&\ln {\cz}=V \Bigl \{\frac{2}{\lambda_N^3}  [ \zeta_n +\zeta_p+
\frac{b_{nn}}{2}\zeta_n^2+\frac{b_{pp}}{2}\zeta_p^2 
+\frac{1}{2}b_{np}\zeta_n\zeta_p  \nonumber \\
&& +8\zeta_\alpha +8b_{\alpha \alpha }\zeta_\alpha^2
+8b_{\alpha n}\zeta_\alpha (\zeta_n+\zeta_p)] \Bigr \},
\eea
where $\lambda_N$=$\frac{h}{\sqrt{2\pi mT}} $ is the nucleon thermal
wavelength, $\zeta_\alpha $=$ e^{\beta (\mu_\alpha +B_\alpha )}$, $B_\alpha $
being the binding energy of $\alpha $ and $\mu_\alpha =2(\mu_n +\mu_p )$.
The $b_{nn}$, $b_{np}$, etc., are the temperature dependent virial coefficients
\cite{hor,mal}. The value of the virial coefficient $b_{np}$ has
been adjusted so as to exclude the resonance formation of deuteron
from n-p scattering to be consistent with our choice of the
$\alpha $N matter.

The knowledge of the partition function allows all the relevant observables
to be calculated. The pressure is given by
\bea
P=T \ln {\cz}/V \,.
\eea
The number density $\rho_i$ is calculated from
\bea
\rho_i=\zeta_i \left (\frac{\partial}{\partial \zeta_i} 
\frac{\ln {\cz}}{V} \right )_{V,T},
\eea
where $i$ stands for n,p, or $\alpha$. Once the pressure,
densities and chemical potentials are known,  the free
energy can be obtained from the Gibbs-Duhem relation. The
entropy per baryon is calculated from
\bea
s_b=\frac{1}{\rho_b}\Bigl (\frac{\partial P}{\partial T} \Bigr )_\mu,
\eea
which yields the energy per baryon as $e_b=f_b+Ts_b$. The explicit
expression for the entropy per baryon is
\bea
&& s_b=\frac{1}{\rho_b}\Biggl \{\frac{5}{2}\frac{P}{T}
-\sum_i \rho_i \ln \zeta_i \nonumber \\
+&& \frac{T}{\lambda_N^3}\bigl [\zeta_n\zeta_p b_{np}^{\prime}+(\zeta_n^2+
\zeta_p^2)b_{nn}^{\prime}  
\nonumber \\
&& +8\zeta_{\alpha}^2b_{\alpha \alpha}^
{\prime}+8\zeta_{\alpha}(\zeta_n+\zeta_p)b_{\alpha n}^{\prime} 
\bigr ]\Biggr \}.
\eea
The prime on the virial coefficients denotes their temperature derivatives.

\section{Results and Discussions}

In the mean-field framework, the momentum and density-dependent 
finite-range modified Seyler-Blanchard force as scripted in 
Eqs.~(1) and (2) has been chosen as the effective two-nucleon interaction
in our calculations. 
To start with, we take baryon matter at a given density $\rho_b $
at a temperature $T$ with an isospin asymmetry $X$. The unknowns 
are the free nucleon densities $\rho_n$, $\rho_p$ and the $\alpha $
concentration in the matter. The three constraints are the conservation
of the total baryon number, the total isospin and the condition of
chemical equilibrium between the nucleons and alphas. 
Starting from a guess value for the
$\alpha $ concentration, the unknowns are determined iteratively
using the Newton-Raphson method. For our calculations, the masses
of neutron and proton are taken to be the same, for $\alpha $
binding energy, the experimental value of 28.3 MeV is used. For
the evaluation of the $\alpha \alpha $-potential, the 
$\alpha $-particles are assumed to be nuclear droplets with
sharp boundary and that they do not interpenetrate.

\begin{figure}
\includegraphics[width=1.0\columnwidth,angle=0,clip=true]{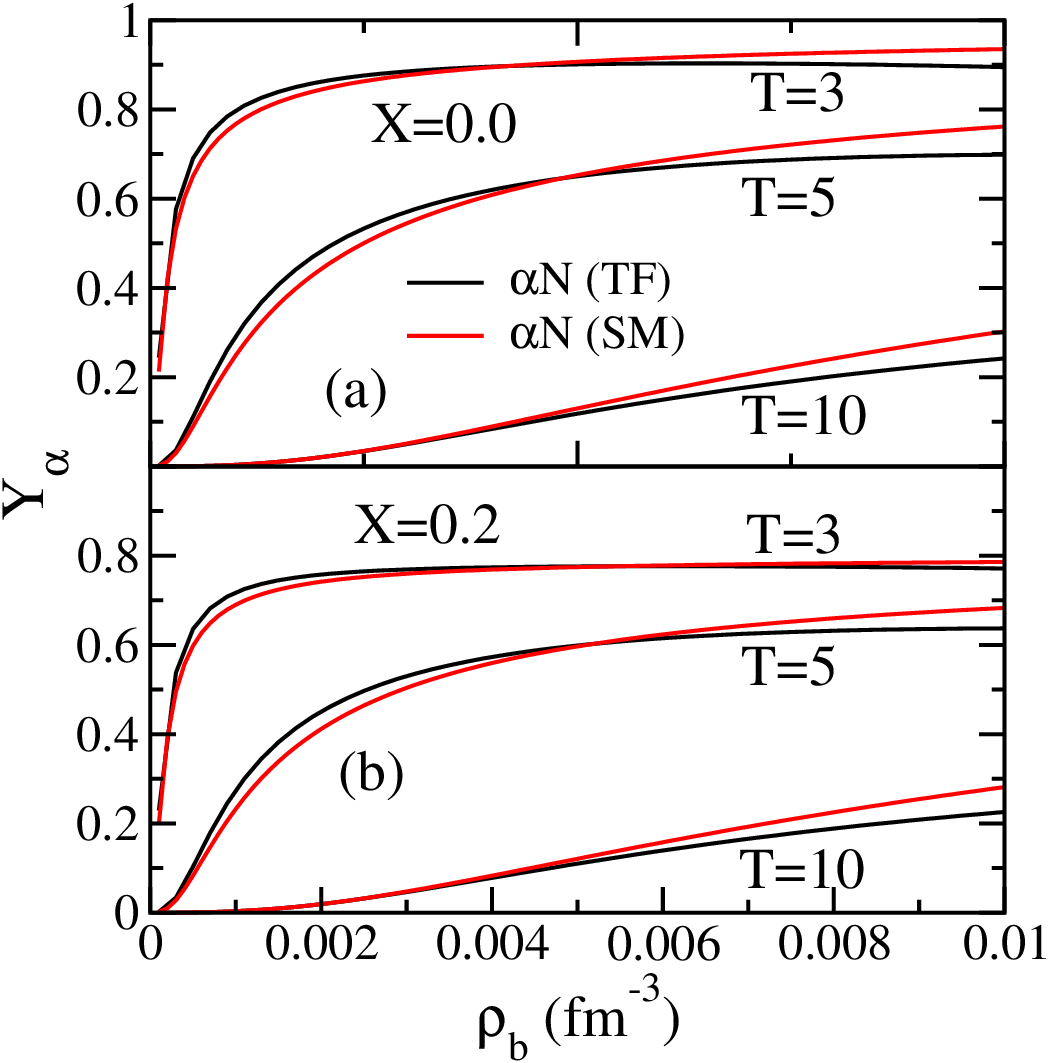}
\caption{(color online) The $\alpha$ fraction $Y_{\alpha}=
4\rho_\alpha /\rho_b $ shown as a function of baryon density
$\rho_b $ in TF and SM approaches at $T$=3, 5 and 10 MeV for
symmetric matter ($X$=0.0) and asymmetric matter ($X$=0.2)
in panels (a) and (b), respectively.} 
\end{figure}

\begin{figure}
\includegraphics[width=1.0\columnwidth,angle=0,clip=true]{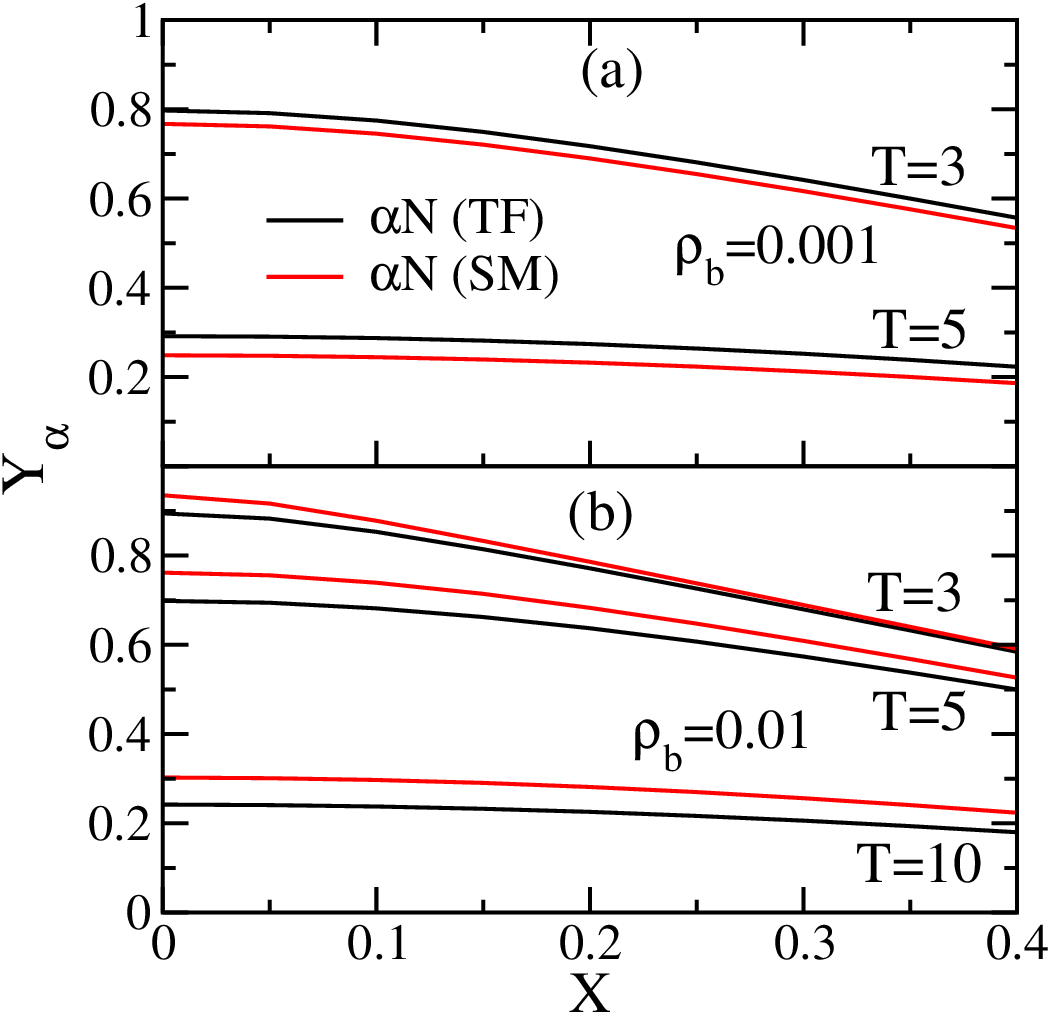}
\caption{(color online)  The $\alpha $ fraction $Y_\alpha $ 
displayed as a function of asymmetry $X$ at baryon density $\rho_b$
=0.001 (upper panel) and at 0.01 fm$^{-3}$ (lower panel) at
$T$= 3, 5 and 10 MeV in TF and SM approaches.}
\end{figure}

\begin{figure}
\includegraphics[width=1.0\columnwidth,angle=0,clip=true]{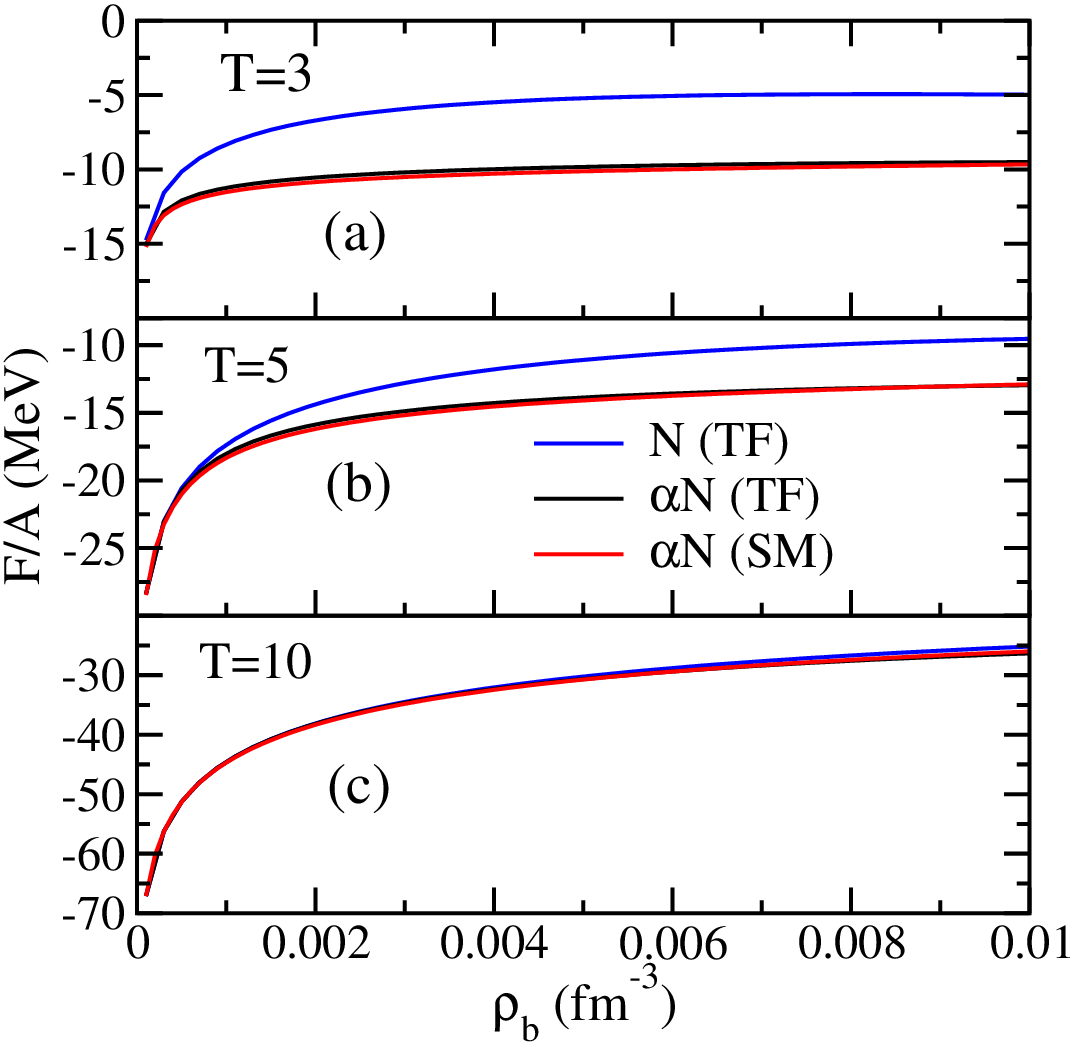}
\caption{(color online) Free energy per baryon $F/A$ shown as a
function of $\rho_b $ at $T$=3, 5 and 10 MeV in the TF framework
for homogeneous nucleonic matter (blue lines) and $\alpha$N
matter (black lines). The red lines represent results from
the SM approach.}
\end{figure}

\begin{figure}
\includegraphics[width=1.0\columnwidth,angle=0,clip=true]{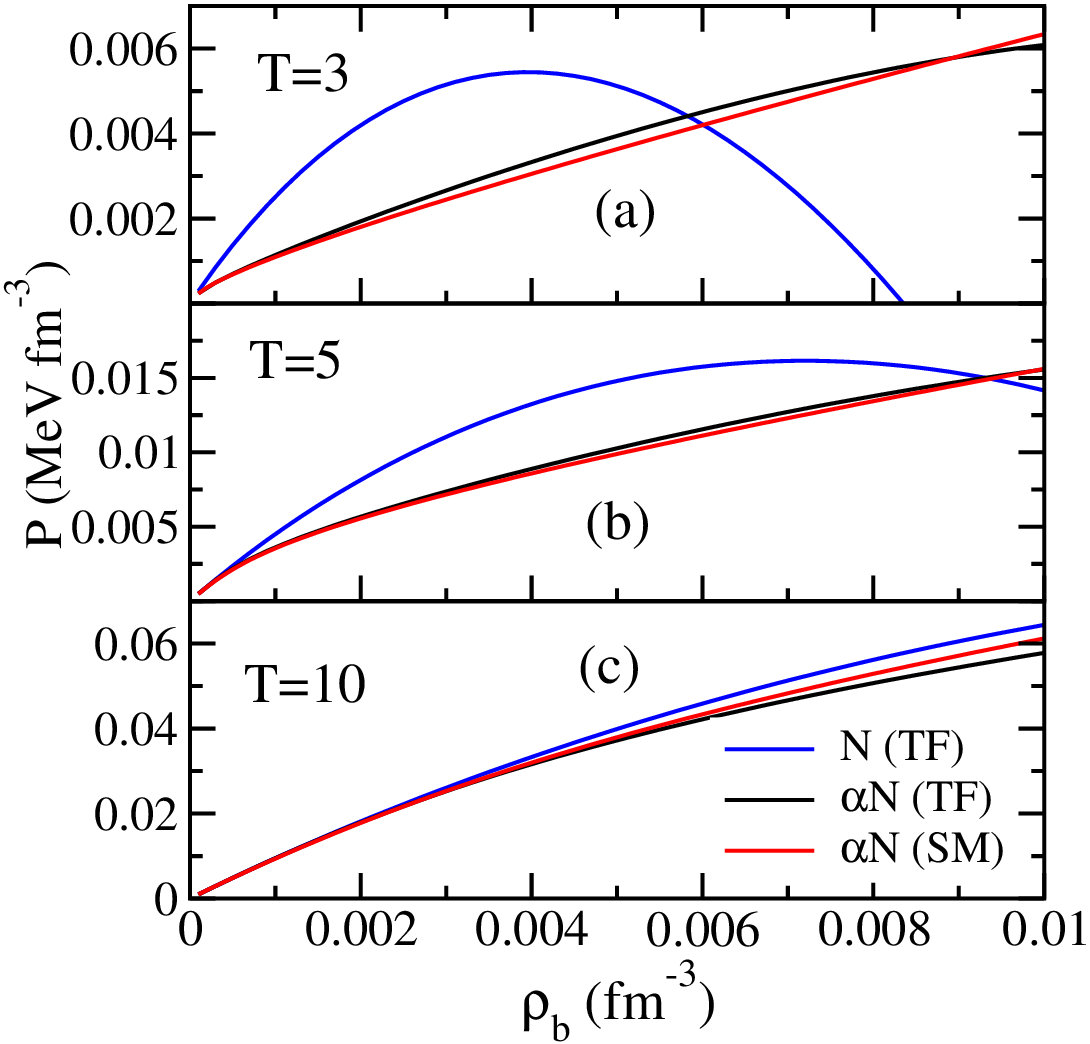}
\caption{(color online) Pressure $P$ as a function of $\rho_b$.
The notations are the same as in Fig.~5.}
 
\end{figure}
\begin{figure}
\includegraphics[width=1.0\columnwidth,angle=0,clip=true]{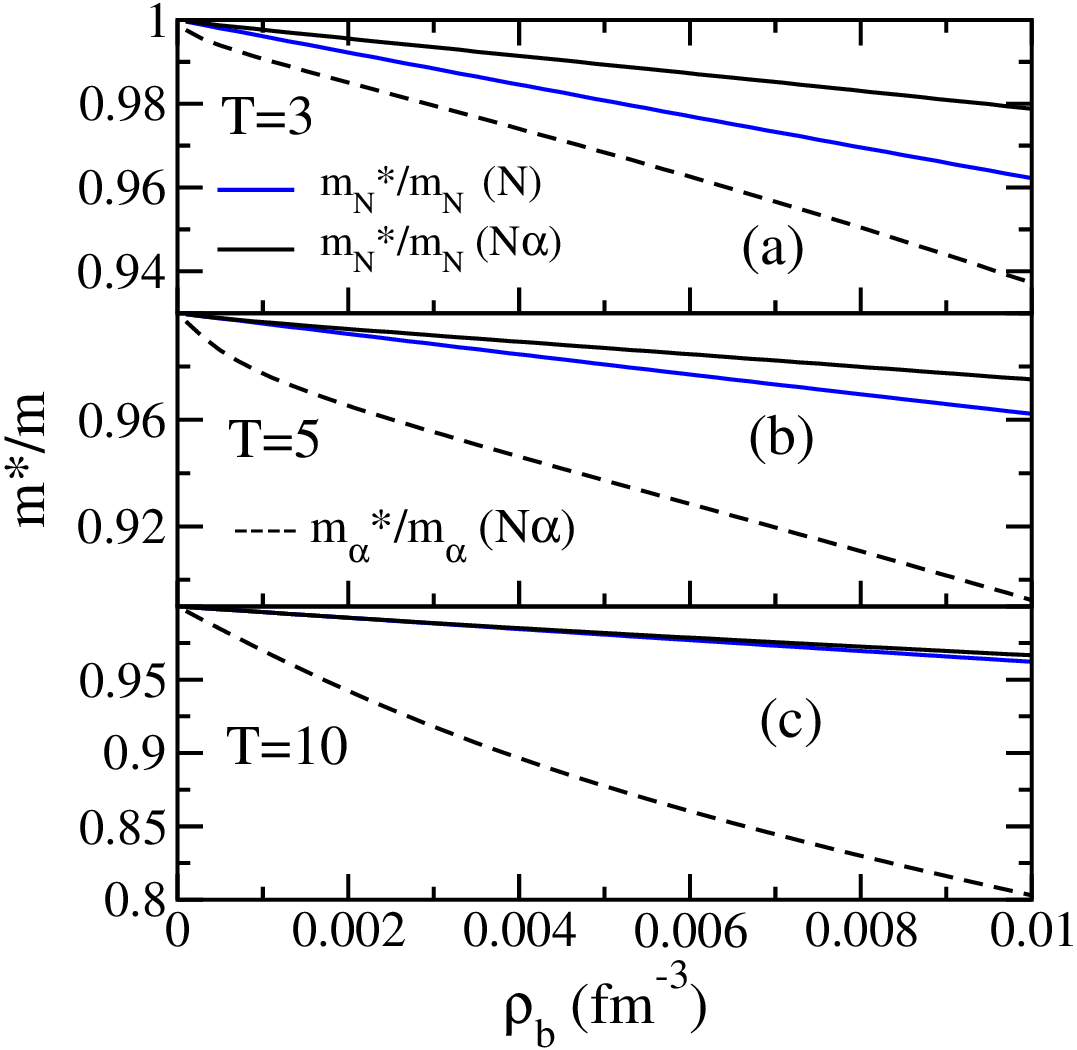}
\caption{(color online) The nucleon (full black lines) and
$\alpha $ (dashed black lines) effective masses shown as a 
function of $\rho_b $ at $T$=3, 5 and 10 MeV in the TF framework
for $\alpha $N matter. The blue lines refer to the corresponding
nucleon effective masses for homogeneous nucleonic matter.}
\end{figure}

The calculations are done upto a baryon density $\rho_b$=0.01 fm$^{-3}$.
To show the effect of temperature on different properties of the dilute
matter, results are reported for temperatures $T$ = 3, 5, and 
10 MeV. In Fig.~3, the baryon fraction in $\alpha $,   
$Y_\alpha $ =$4\rho_\alpha /\rho_b $ (hereafter referred to as 
$\alpha $ fraction) in $\alpha $N matter 
 as a function of density at 
the three temperatures mentioned are shown for symmetric ($X$=0) and
asymmetric ($X$ =0.2) nuclear matter in panels (a) and (b),
respectively. The black lines correspond to results obtained 
in the TF approximation [$\alpha $N (TF)], the red lines refer to
those in the SM approach [$\alpha $N (SM)] with consideration of
only n, p and $\alpha $ as the constituents of the baryonic matter. 
At low temperatures and higher densities, it is seen 
that alphas are the major constituents of the matter, with 
increasing temperature, the free nucleon fraction increases at the 
cost of $\alpha $ density. At moderate asymmetry $X$=0.2,
the $\alpha $ population is somewhat lower compared to that for
symmetric nuclear matter. In the temperature and density domain 
that we explore, the results from both 
the SM and  TF approach are found to be quite close. 
The asymmetry dependence of $\alpha $ fraction $Y_\alpha $ is displayed
in Fig.~4 at two representative densities $\rho_b$=0.001 and
0.01 fm$^{-3}$ at the three temperatures. With increasing 
asymmetry, the $\alpha $ concentration decreases, the decrease is 
more prominent at lower temperature. At the lower density (Fig.~4(a)),
results for $T$=10 MeV are not shown as $Y_\alpha $ is close to zero.

In Fig.~5, the free energy per baryon for the homogeneous nucleonic
matter (denoted by N(TF)) and the $\alpha $N matter in the TF
approximation are presented in panels (a), (b), and (c) at $T$=
3, 5, and 10 MeV, respectively. The calculations presented refer
to symmetric nuclear matter. The blue and black lines represent
results for N(TF) and $\alpha $N(TF). It is clearly
seen that the clusterized matter has lower free energy compared to 
homogeneous nucleonic matter. This is more prominent at lower temperatures,
higher temperature tends to melt away the clusters. For comparison,
results from the $S$-matrix approach are also presented. They are
shown by the red lines, nearly indistinguishable from those
from $\alpha $N(TF). Fig.~6
displays the pressure of the baryonic matter.
At lower temperatures ($T$=3 and 5 MeV), the nucleonic matter
shows the rise and fall of the pressure with density leading to
unphysical region. For $\alpha $N matter, however, no such 
unphysical region is observed in the density region 
we have studied. 
Both the TF and the SM approaches 
yield nearly the same value of pressure. At high temperature
the $\alpha $ concentration becomes very less, 
the pressure in all the three approaches  are then nearly the same 
in this density region.

In Fig.~7, the effective masses of  nucleon and $\alpha $ 
are shown as a function of density at the  temperatures mentioned. 
The nucleon effective mass
is calculated for both nucleonic matter (blue line) and $\alpha $N
matter (full black line) in the TF approximation. 
The nucleon effective mass at a given $\rho_b $ in homogeneous
nucleonic matter is always lower compared to that in clusterized 
matter. It is independent of temperature. In $\alpha $N
matter it nominally decreases with temperature.
At high temperature,
the nucleon effective masses calculated in the homogeneous and clusterized
matter are nearly degenerate, with lowering of temperature, the 
degeneracy is lifted due to the increase in the  $\alpha $ concentration.
The effective $\alpha $ mass is shown by the dashed black lines.
With increasing temperature, the medium effect on the $\alpha $ mass
gets strikingly enhanced. This is due to the interplay of the 
temperature-dependent contributions from the $\alpha \alpha $ interactions
and $\alpha $N interactions corresponding to the first and the second
term within the braces in Eq.~(29).

\begin{figure}
\includegraphics[width=1.0\columnwidth,angle=0,clip=true]{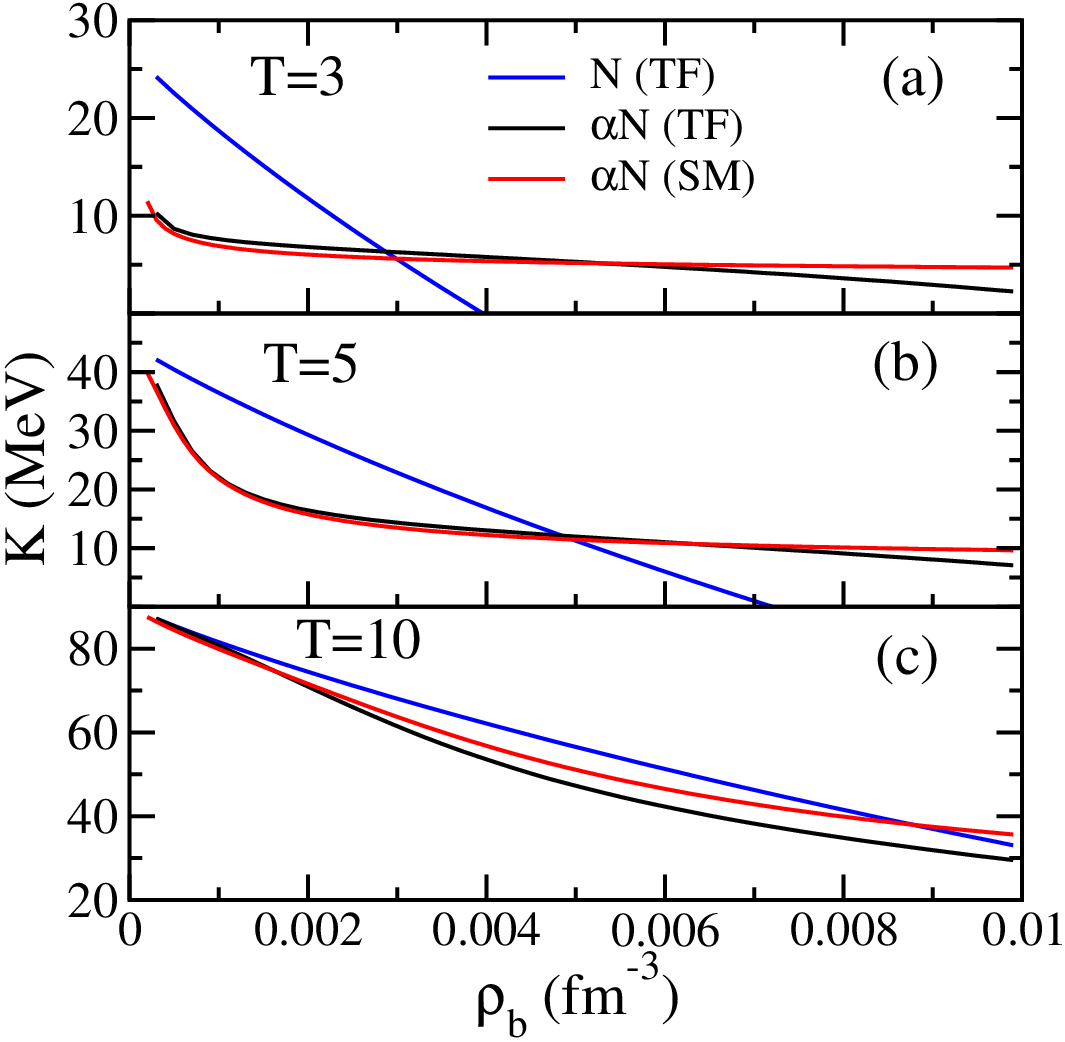}
\caption{(color online) The incompressibility $K$ for baryonic matter
shown as a function of $\rho_b$ at $T$=3, 5 and 10 MeV. The
notations are the same as in Fig.~5.}
\end{figure}

The incompressibility of the baryonic matter as a function of density
is displayed in Fig.~8 at the three temperatures. 
At very low density and higher temperature, the matter is mostly nucleonic
in all the three approaches, so the incompressibility  $K$ is $\sim $
9$T$; this one sees at the lower densities considered at 
$T$ =10 MeV in panel (c) of this figure.
Even at this very high temperature, however, the nucleonic
interactions have their role as the density increases;
this results in the reduction of the incompressibility from the
ideal gas value. 
 At the lower temperatures (panels (a) and (b)),
clusterization softens the matter towards compression compared to
homogeneous matter (shown in the lower density region); increasing
density, however, pushes the homogeneous matter towards 
 the unphysical region leading to negative
incompressibility.
\begin{figure}
\includegraphics[width=1.0\columnwidth,angle=0,clip=true]{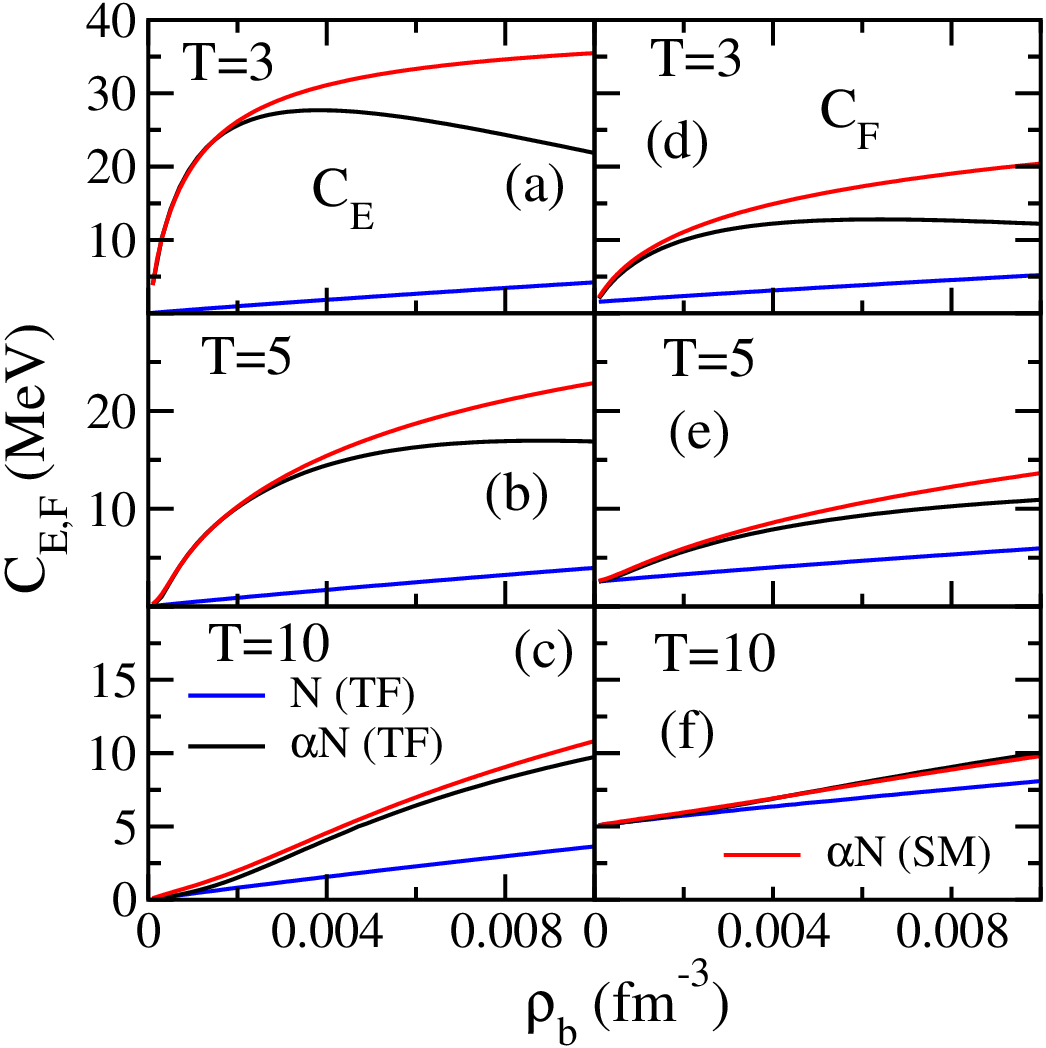}
\caption{(color online)  The symmetry energy $C_E$ (left panels)
and symmetry free energy coefficients $C_F$ (right panels) shown
as a function of $\rho_b$ at temperatures $T$=3, 5 and 10 MeV.
The notations are the same as in Fig.~5.}
\end{figure}

\begin{figure}
\includegraphics[width=1.0\columnwidth,angle=0,clip=true]{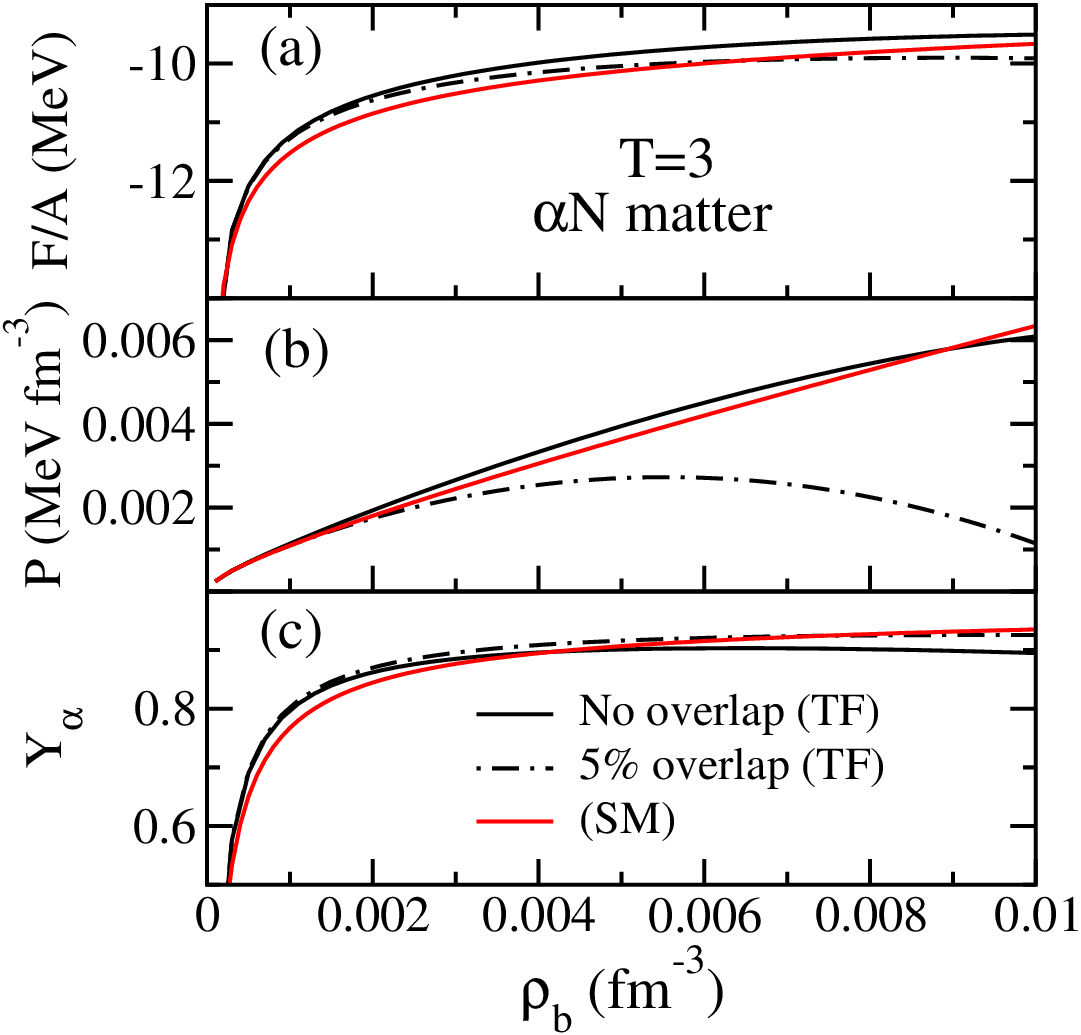}
\caption{(color online)  The free energy per particle, pressure
and $\alpha $ fraction  shown as a function of $\rho_b$ at
$T$=3 MeV in panels (a), (b) and (c), respectively, for $\alpha$
drops with no overlap (full black lines) and with at best 5$\% $
overlap (dashed-dot black lines) in the TF approximation. The
same observables are also shown in the SM approach (red lines).}
\end{figure}

The symmetry energy coefficients $C_E$ and $C_F$ of the baryonic
matter as a function of density are displayed in the left and 
right panels, respectively, of Fig.~9 at the three temperatures
studied. The blue lines refer to calculations for the homogeneous
matter, the black and red lines represent results for $\alpha $N(TF)
and $\alpha $N(SM). Clusterized matter displays a marked increase
in the symmetry coefficients noticed already earlier \cite{hor,sam}.
The two approaches to clusterization  
lead to the same values of the symmetry 
coefficients at lower densities, with increase in density
the difference widens, more so at lower temperatures.

 The results presented so far in the $\alpha $N(TF) approach have been
calculated with the assumption that the alphas do not overlap, they are
mutually impenetrable spherical drops. This assumption
relies on the fact that the alphas are very tightly bound and very hard
to excite.
To explore the effect of overlap in alphas, we consider a possibility
of penetration with at best a 5 $\% $ overlap in volume 
(the value of $I_\alpha $
in Eq.~(30) then changes accordingly). Calculations have been repeated
with this changed  condition. The so-calculated free energy per baryon,
pressure and the $\alpha $ fraction $Y_\alpha $ in the baryonic matter
are presented in panels (a), (b) and (c), respectively,
of Fig.~10 at $T$ =3 MeV
(the dot-dashed black lines) and compared with those calculated with the 
no-overlap condition (the full black lines) and also those from the 
$\alpha $N(SM) approach (the red lines). There is no significant change 
in the free energy or in $\alpha $ fraction, but the pressure changes 
perceptibly, particularly at higher density. The good agreement
between the no-overlap $\alpha $N(TF) calculations with those from the
bench-mark $\alpha $N(SM) shows the viability 
of the approximation of the impenetrability
of the alphas.

\section{Concluding remarks}

Clusterization in warm dilute nuclear matter has been treated 
earlier in the virial approach or in the $S$-matrix framework. These
are model-independent parameter-free calculations. 
As explained in the introduction, these methods may have limitations
at relatively high densities and low temperatures. An  alternate 
avenue for dealing with clusterized matter in a broadened density
and temperature domain is suggested in the mean-field framework
in the present paper. The suggested method may be lengthy at
relatively higher densities where many different fragment species
are formed, but it is straightforward. To explore its applicability
in a wider domain, as a first step,
we consider only
n, p and $\alpha $ as the constituents of the matter
at low densities and see how
the results compare with those from the model-independent virial
approach.

We have chosen the SBM interaction that 
nicely reproduces the bulk properties
of nuclear matter and of finite nuclei. We have calculated
the $\alpha $ fraction, free energy, pressure, incompressibility and
the symmetry coefficients of this $\alpha $N matter in this mean-field
framework and find that all these results compare extremely well with
those obtained from the $S$-matrix method, particularly in the 
low-density high-temperature regime.
This gives one confidence
in the applicability of this mean-field approach in dealing with the
EOS of warm dilute baryonic matter and the possibility of extending
this method to higher densities. 
The price, however, is consideration of a larger number of fragment
species and a numerically involved calculation.

\begin{acknowledgments}
  S.K.S. and J.N.D acknowledge support of DST, Government of India.
\end{acknowledgments}


\begin{thebibliography}{99}
\bm{cla} J. W. Clark and T. P. Wang, Ann. Phys. (N.Y.), {\bf 40},
127 (1966).

\bm{car} F. Carstoiu and S. Misicu, Phys. Lett. B{\bf 682},
33 (2009).

\bm{lamb} D. Q. Lamb, J. M. Lattimer, C. J. Pethick, and D. G.
Ravenhall, Phys. Rev. Lett. {\bf 41}, 1623 (1978).

\bibitem{fri} B. Friedman and V. R. Pandharipande, Nucl. Phys. 
A {\bf 361}, 502 (1981).

\bibitem{pei} G. Peilert, J. Randrup, H. Stocker, and W. Greiner,
Phys. Lett. B {\bf 260}, 271 (1991).

\bibitem{sam} S. K. Samaddar, J. N. De, X. Vi\~nas, and M. Centelles,
Phys. Rev. C {\bf 80}, 035803 (2009). 

\bibitem{hor} C. J. Horowitz and A. Schwenk, Nucl. Phys. A {\bf 776},
55 (2006).

\bibitem{de} J. N. De, S. K. Samaddar, and B. K. Agrawal, 
Phys. Rev. C {\bf 82}, 045201 (2010).

\bibitem{jan} H.-Th. Janka, K. Langanke, A. Marek, G. Mart\'inez-Pinedo,
and B. M\"uller, Phys. Rep. {\bf 442}, 38 (2007).

\bibitem{fuc} C. Fuchs, J. Phys.G {\bf 35}, 014049 (2008).


\bibitem{con} E. O'connor, D. Gazit, C. J. Horowitz, A. Schwenk,
and N. Barnea, Phys. Rev. C {\bf 75}, 055803 (2007).

\bm{mal} S. Mallik, J. N. De, S. K. Samaddar, and Sourav Sarkar, 
Phys. Rev. C {\bf 77}, 032201 (R) (2008).

\bm{pai} A. Pais and G. E. Uhlenbeck, Phys. Rev. {\bf 16},
250 (1959).

\bm{de1} J. N. De, N. Rudra, Subrata Pal, and S. K. Samaddar,
Phys. Rev. C {\bf 53}, 783 (1996).

\bm{uma} V. S. Uma Maheswari, D. N. Basu, J. N. De and S. K. Samaddar,
Nucl. Phys. {bf A615}, 516 (1997).

\bm{mye} W. D. Myers and W. J. Swiatecki, Ann. Phys. (N.Y.) {\bf 204},
401 (1990).

\bm{das} R. Dashen, S-k. Ma and H.J. Bernstein, Phys. Rev, 187, 345 (1969).


\end{thebibliography}
\end{document}